\documentclass[aps,prb,twocolumn,showpacs,amssymb]{revtex4}
\usepackage{graphicx}
\usepackage{bm}

\begin{document}

\title{$g$-factors and discrete energy level velocities in nanoparticles}

\author{Eduardo R. Mucciolo,$^1$ Caio H. Lewenkopf,$^2$ and Leonid
I. Glazman$^3$}

\affiliation{$^1$Department of Physics, University of Central Florida,
P.O. Box 162385, Orlando, Florida 32816-2385, USA}

\affiliation{$^2$Instituto de F\'{\i}sica, Universidade do Estado do
Rio de Janeiro, R. S\~ao Francisco Xavier 524, 20550-900 Rio de
Janeiro, Brazil,}

\affiliation{$^3$Theoretical Physics Institute, University of
Minnesota, Minneapolis, Minnesota 55455, USA}

\date{\today}

\begin{abstract}
We establish relations between the statistics of $g$ factors and the
fluctuations of energy in metallic nanoparticles where spin-orbit
coupling is present. These relations assume that the electron dynamics
in the grain is chaotic. The expressions we provide connect the second
moment of the $g$ factor to the root-mean square ``level velocity''
(the derivative of the energy with respect to magnetic field)
calculated at magnetic fields larger than a characteristic correlation
field. Our predictions relate readily observable quantities and allow
for a parameter-free comparison with experiments.
\end{abstract}


\maketitle


It was noted in experiments \cite{salinas99,davidovic99} that the
Zeeman splitting of discrete energy levels in nanoparticles is very
sensitive to the presence of spin-orbit interaction. The splitting can
be characterized by a level-dependent $g$ factor. Adding to the Al
grains only 4\% of Au resulted in a change of measured $g$ factors
from approximately $1.7$ to $0.7$. In addition to being suppressed,
the $g$ factor in the presence of spin-orbit interaction also
fluctuates randomly from level to level.

These fluctuations of $g$ factors were described in the framework of
random matrix theory \cite{brouwer00,matveev00,serota01} (RMT) and the
suppression of the $g$ factor was related to the strength of the
spin-orbit interaction and to the elastic mean free path of electrons
in the grains.\cite{matveev00} The limit of strong spin-orbit
interaction corresponds to a short spin-orbit scattering time,
$\tau_{\rm so}\delta/\hbar \ll 1$, where $\delta^{-1}$ is the mean
density of states in the grain at zero field. It was predicted that
the distribution of eigenvalues of the $g$-factor tensor in this limit
should have a Gaussian form. Within RMT, this distribution is
characterized by a phenomenological parameter $\left\langle g^2
\right\rangle$. In Ref. \onlinecite{matveev00}, this parameter was
expressed in terms of the grain size, electron mean free path $l$, and
relaxation time $\tau_{\rm so}$.
However, a comparison of the experimental results with theory was not
entirely satisfactory. On the one hand, there is an indication that
the distribution of the eigenvalues and directions of the eigenvectors
of the $g$-tensor is Gaussian and correspond to a ``pure'' symplectic
ensemble.\cite{petta01,petta02} On the other hand, it is not clear
\cite{petta01} whether the small values of $\left\langle g^2
\right\rangle$ obtained in the experiments agree with the theoretical
values estimated in Ref. \onlinecite{matveev00}. The difficulty in
making the comparison comes from the lack of information about the
amount of disorder in the grain.

The goal of this work is to provide three relations for the
distribution width $\left\langle g^2 \right\rangle$ to other
quantities that are directly measured in the same set of experiments
(and thus do not rely on any additional information about the amount
of disorder in the grains). These quantities are the variance of the
energy level derivative with respect to the magnetic field (known as
level velocity \cite{szafer93}) and the zero-magnetic field level
curvature (the second derivative of the energy level with respect to
magnetic field).


We begin by stating our main results. Our first expression, valid for
strong spin-orbit coupling only, $\tau_{\rm so}\, \delta/\hbar \ll 1$,
is
\begin{equation}
\label{eq:g2}
\left\langle g^2 \right\rangle = \frac{12}{\mu_B^2} {\rm var} \left[
\left( \frac{d\varepsilon_\nu}{dB} \right)_{B\gg B^\ast}
\right]_{\tau_{\rm so}\rightarrow 0},
\end{equation}
where var[\,] denotes the variance, $\mu_B$ is the Bohr
magneton, and $B^\ast$ is the crossover field for breaking
time-reversal symmetry. Equation (\ref{eq:g2}) gives a statistical
connection between the response at $B\to 0$ (the $g$ factor) to that
at large magnetic fields (the level velocity). It provides a way to
check experimentally if the grains exhibiting fluctuations of the
$g$ factor indeed belong to a ``pure'' symplectic ensemble, rather
than to an ensemble describing the crossover between the orthogonal
and symplectic limits. Provided that one collects data for the
dispersion of energy levels over a sufficiently large range of
magnetic fields, the terms on both sides of Eq. (\ref{eq:g2}) can be
found independently using the same data set. This expression is
universal and contains no microscopic or materials parameters.

The second expression we find relates properties of two sets of grains
which are equivalent macroscopically except for the value of
$\tau_{\rm so}$. It reads
\begin{equation}
\label{eq:g2_mix}
\left\langle g^2 \right\rangle = \frac{3\, g_0^2}{2\pi\hbar} \tau_{\rm
so} \delta + \frac{3}{2\, \mu_B^2} {\rm var} \left[
\sum_{\sigma=\uparrow,\downarrow} \left(
\frac{d\varepsilon_{n\sigma}}{dB} \right)_{B\gg B^\ast}
\right]_{\tau_{\rm so} \rightarrow \infty},
\end{equation}
where $g_0$ denotes the materials bulk value for the $g$ factor
($g_0=2$ for free electrons). In Eq. (\ref{eq:g2_mix}), $\left\langle
g^2 \right\rangle$ is evaluated in the strong-spin-orbit coupling
limit ($\tau_{\rm so} \rightarrow 0$). However, the second term on the
right-hand side is evaluated for $\tau_{\rm so} \rightarrow \infty$
(absence of spin-orbit coupling). For example, one may consider two
sets of Al:Au grains, one with no doping and another with moderate
doping.\cite{salinas99}

The two terms on the right-hand side of Eq.~(\ref{eq:g2_mix}) come
from distinct contributions. The first term, the ``spin part,'' is
associated with the Debye mechanism of energy dissipation associated
with spin reorientations; the second one, the ``orbital part,'' is due
to the eddy currents induced in the grain. We note that the spin-orbit
scattering rate $1/\tau_{\rm so}$ apparently may be estimated for a
given material or host-dopant pair.\cite{petta01} Equation
(\ref{eq:g2_mix}) permits us to separate the spin and orbital
contributions to the fluctuations of the $g$ factor.

The second moment of the $g$ factor can also be related to the
statistics of other spectral quantities, such as the zero-field level
curvature
\begin{equation}
\label{eq:gK}
\left\langle g^2 \right\rangle = \frac{9\, \delta}{2\, \sqrt{2}\,
\mu_B^2} \left\langle \left| \left( \frac{d^2\varepsilon_\nu}{dB^2}
\right)_{B=0} \right| \right\rangle.
\end{equation}
As Eq. (\ref{eq:g2}), this expression is applicable in the
strong-spin-orbit coupling regime only (symplectic ensemble).

It is important to remark that the $g$ factor in a given metallic
nanoparticle is in reality a tensor with three different eigenvalues,
even for grains that are statistically
isotropic.\cite{brouwer00,petta02} However, the distribution of the
matrix elements of the $g$-factor tensor is still characterized by a
single quantity $\left\langle g^2 \right\rangle$. Throughout our
manuscript, when establishing relations between $\left\langle g^2
\right\rangle$ to other quantities, we consider a magnetic field
applied in some fixed but arbitrary direction. Once these relations
are established, one may use Ref. \onlinecite{brouwer00} to construct
the full statistics of the $g$-factor tensor.


We will now establish Eqs. (\ref{eq:g2})-(\ref{eq:gK}). The main idea
behind the derivations is to relate statistical quantities to
invariants of the system, such as the traces of the magnetic moment
operator. For that purpose, let us begin by writing the Hamiltonian
for the disordered (or chaotic) grain in the presence of an applied
magnetic field as $\hat{H}(B) = \hat{H}_0 + B\, \hat{M}$, where the
magnetic moment operator has both orbital and spin parts: $\hat{M} =
\hat{M}_{\rm orb} + \hat{M}_{\rm spin}$. To simplify the discussion,
we assume that the grain is isotropic. We define the $g$ factor of the
$n$th energy level as
\begin{equation}
\label{eq:gn}
g_n \equiv \frac{1}{2\mu_B}\, \left| \left(
\frac{d\varepsilon_{n\sigma}}{dB} \right)_{B=0} \right|,
\end{equation}
where $\{\varepsilon_{n\sigma}\}$ are the eigenvalues of
$\hat{H}_0$. Note that due to Kramers degeneracy at $B=0$, the levels
are twofold degenerate. We use the index $\sigma$ to distinguish
states that are time-reversal conjugate.  The matrix elements of
$\hat{H}_0$ follow either the symplectic ($\beta=4$) or orthogonal
($\beta=1$) ensemble statistics, depending on whether spin-orbit
coupling is present or absent, respectively. For both cases, the
matrix elements of $\hat{M}$, when expressed in the eigenbasis
$\{|n\sigma; 0\rangle \}$ of $\hat{H}_0$, fluctuate according to a
Gaussian distribution with zero mean. For the symplectic ensemble, the
variance of the diagonal matrix elements reads\cite{matveev00}
\begin{equation}
\label{eq:Melem}
\left\langle |\langle n \sigma; 0 | \hat{M} | n \sigma; 0 \rangle|^2
\right\rangle_{\beta=4} = \frac{3\, {\rm Tr} \left( \hat{M}^2
\right)}{4N^2},
\end{equation}
where the trace runs over the $2N$ states in the band, with $N \gg 1$
being assumed (the factor of 2 accounts for Kramers degeneracy). The
degeneracy at zero field allows us to pick a basis such that $\langle
n\sigma;0 |\hat{M}| n \sigma^\prime; 0 \rangle$ is diagonal in the
$\sigma$ indices. Using Eq. (\ref{eq:gn}) and first-order perturbation
theory, we find that $\langle n\sigma;0 |\hat{M}| n \sigma^\prime; 0
\rangle = (-1)^\sigma\, \delta_{\sigma\sigma^\prime}\, g_n\,
\mu_B/2$. Thus, from Eq. (\ref{eq:Melem}), we arrive
at\cite{zirnbauer92}
\begin{equation}
\label{eq:g2M}
\left\langle g^2 \right\rangle_{\beta = 4} = \frac{3}{\mu_B^2}\,
\frac{{\rm Tr}\left( \hat{M}^2 \right)} {N^2}.
\end{equation}
Notice that the quantities on the left-hand side of
Eqs. (\ref{eq:Melem}) and (\ref{eq:g2M}) are defined at $B=0$.


Now we have to write the statistical quantities that appear on the
right-hand side of Eqs. (\ref{eq:g2})-(\ref{eq:gK}), in terms of ${\rm
Tr} ( \hat{M}^2)$. Note that the latter is an invariant and therefore
takes the same value at zero or large magnetic fields.

Let us first consider Eq. (\ref{eq:g2}). The variance of the level
velocity can be computed in terms of the variance of the matrix
elements of the magnetic moment operator since
\begin{equation}
\label{eq:velM}
\left( \frac{d\varepsilon_\nu}{dB} \right)_{B=B_0} = \langle \nu; B_0
| \hat{M} | \nu; B_0 \rangle.
\end{equation}
For a sufficiently large magnetic field $B_0 \gg B^\ast$,
time-reversal symmetry in the grain is broken. In the presence of
strong-spin-orbit scattering, orbital and spin degrees of freedom
remain mixed, but the ensemble statistics of the Hamiltonian
eigenstates switches from $2N \times 2N$ symplectic to $2N \times 2N$
unitary ($\beta=2$). Thus, we need to compute the variance of the
matrix elements of the magnetic moment operator in the unitary
regime. For this purpose, we make use of the eigenvalues and
eigenvectors of the magnetic moment operator: $\hat{M}|k\alpha\rangle
= (-1)^\alpha M_k\, |k\alpha\rangle$, with $k=1,\ldots,N$ and $\alpha
= \pm 1$ due to the time-reversal properties of $\hat{M}$.\cite{obs2}
This yields
\begin{equation}
\label{eq:MelemB0}
\langle \nu; B_0 | \hat{M} | \nu; B_0 \rangle = \sum_{k,\alpha}
(-1)^\alpha M_k\, |\langle \nu; B_0 | k\alpha \rangle|^2.
\end{equation}
For the unitary ensemble in the large-$N$ limit, the eigenvector
amplitudes shown on the right-hand side of Eq. (\ref{eq:MelemB0})
fluctuate independently according to the Porter-Thomas
distribution.\cite{porterthomas} One finds that $\left|\langle \nu;
B_0 | k\alpha \rangle|^2 \right\rangle = \frac{1}{2} N$ and
$\left\langle |\langle \nu; B_0 | k\alpha \rangle|^4 \right\rangle =
\frac{1}{2} N^2$, independently of state indices. Hence, the average
matrix element of $\hat{M}$ must vanish and the variance can be
written as
\begin{equation}
\label{eq:varM}
{\rm var} \left[ \langle \nu; B_0 | \hat{M} | \nu; B_0
\rangle_{\tau_{\rm so} \rightarrow 0,\, \beta=2} \right] = \frac{{\rm
Tr} \left( \hat{M}^2 \right)} {4N^2}.
\end{equation}
Putting together Eqs. (\ref{eq:g2M}), (\ref{eq:velM}), and
(\ref{eq:varM}), we arrive at Eq. (\ref{eq:g2}).


To derive Eq. (\ref{eq:g2_mix}), we separate the magnetic moment in
terms of spin and orbital parts, $\hat{M} = \hat{M}_{\rm spin} +
\hat{M}_{\rm orb}$, which are statistically independent from each
other. From Eq. (\ref{eq:g2M}), we obtain
\begin{equation}
\label{absorption}
\left\langle g^2 \right\rangle_{\beta = 4} = \frac{3}{\mu_B^2\, N^2}\,
\left[ {\rm Tr} \left( \hat{M}_{\rm spin}^2 \right) + {\rm Tr} \left(
\hat{M}_{\rm orb}^2 \right) \right].
\end{equation}
The spin contribution can be written in terms of the imaginary part of
the ac spin susceptibility of a free electron gas in the presence of
spin-orbit coupling.\cite{matveev00} The susceptibility can then be
evaluated using conventional means of rate equations at frequencies
much larger than the mean level spacing in the grain, yet smaller than
the spin-orbit scattering rate. In the limit of $\tau_{\rm so}
\delta/\hbar \ll 1$ one finds\cite{matveev00}
\begin{equation}
\frac{{\rm Tr} \left( \hat{M}_{\rm spin}^2 \right)} {N^2} =
\frac{g_0^2\, \mu_B^2}{2\pi\hbar}\, \tau_{\rm so} \delta.
\end{equation}
This corresponds to the first term on the right-hand side of
Eq. (\ref{eq:g2_mix}).

The orbital contribution in Eq. (\ref{absorption}) is ensemble
independent and is the same regardless of the presence or absence of
the spin-orbit interaction. It is convenient to evaluate it in the
limit of $\tau_{\rm so} \rightarrow \infty$ to decouple spin and
orbital degrees of freedom. To relate the ${\rm Tr} \left(
\hat{M}_{\rm orb}^2 \right)$ to the variance of the velocity of
spin-resolved levels at large fields, we can use Eq. (\ref{eq:velM})
and note that $\hat{M}_{\rm spin}$ just induces a constant slope in
the dispersion of the energy levels with magnetic field at $\tau_{\rm
so} \rightarrow \infty$, while all fluctuations are caused by
$\hat{M}_{\rm orb}$. Therefore, we write
\begin{equation}
\label{eq:velM2}
\sum_{\sigma = \uparrow,\downarrow} \left(
\frac{d\varepsilon_{n\sigma}}{dB} \right)_{B=B_0} = \langle n\uparrow;
B_0 |\hat{M}_{\rm orb}| n\uparrow; B_0 \rangle
\end{equation}
to isolate the fluctuating part (note that $B_0 \gg B^\ast$ here as
well). It is important to observe that in the absence of spin-orbit
mixing and for large magnetic fields the statistics of the eigenstate
$|n\uparrow;B_0\rangle$ corresponds to a $N\times N$ unitary ensemble
(rather than to $2N \times 2N$ when spin and orbital parts are
strongly coupled). Similarly to Eq. (\ref{eq:MelemB0}), we can
decompose the matrix elements in Eq. (\ref{eq:velM2}) using the
eigenstates of $\hat{M}_{\rm orb}$. In this case, however, due to the
change in the ensemble dimension, we have $\left\langle |\langle
n\uparrow; B_0 | k\lambda \rangle|^2 \right\rangle = 1/N$ and
$\left\langle |\langle n\uparrow; B_0 | k\lambda \rangle|^4
\right\rangle = 2/N^2$ for large $N$, independently of the orbital
quantum number $n$. Finally, combining all these results, we arrive at
\begin{equation}
\label{eq:varvelO}
{\rm var} \left[ \sum_{\sigma=\uparrow,\downarrow} \left(
\frac{d\varepsilon_{n\sigma}}{dB} \right)_{B=B_0} \right]_{\tau_{\rm
so} \rightarrow \infty} = \frac{2\, {\rm Tr} \left( \hat{M}_{\rm orb}
\right)^2}{N^2}.
\end{equation}
Inserting this expression into Eq. (\ref{absorption}), we obtain the
second term on the right-hand side of (\ref{eq:g2_mix}). We remark
that Eq.  (\ref{eq:g2_mix}) is fully consistent with Eq. (31) of
Ref. \onlinecite{adam02}, where a description of the statistical
properties of the $g$ factor including the intermediate crossover
regime was developed in terms of phenomenological RMT parameters.


In order to obtain Eq. (\ref{eq:gK}) we follow an approach similar to
that employed in the two previous derivations. We define the level
curvature at $B=0$ as
\begin{equation}
\label{eq:Kdef}
K_n \equiv \left( \frac{d^2 \varepsilon_{n\sigma}} {dB^2}
\right)_{B=0} = 2\sum_{n^\prime\neq n} \frac{ \sum_{\sigma^\prime}
|\langle n \sigma; 0 | \hat{M} | n^\prime \sigma^\prime; 0 \rangle|^2}
{\varepsilon_n - \varepsilon_{n^\prime}}.
\end{equation}
(To simplify the notation, here we set $\varepsilon_{n\sigma} =
\varepsilon_n$.)  Since eigenvalues and eigenfunctions fluctuate
independently in the Gaussian ensembles, we find that
\begin{eqnarray}
\label{eq:K2GSE}
\left\langle K_n^2 \right\rangle_{\beta=4} & = & \frac{8}{\Delta^2}
\left[ \left\langle |\langle n\sigma;0 |\hat{M}| n^\prime
\sigma^\prime;0 \rangle|^4 \right\rangle \right. \nonumber \\ & &
\left. + \left\langle |\langle n\sigma;0 |\hat{M}| n^\prime
\sigma^\prime;0 \rangle |^2 \right\rangle^2 \right],
\end{eqnarray}
with $n\neq n^\prime$ and $\sigma, \sigma^\prime$ taking arbitrary
values. The prefactor in Eq. (\ref{eq:K2GSE}) is defined as
\begin{equation}
\label{eq:Delta2}
\frac{1}{\Delta^2} \equiv 2\delta \sum_n \sum_{n^\prime\neq n}
\left\langle \frac{\delta ( \varepsilon_n )}{( \varepsilon_n -
\varepsilon_{n^\prime})^2} \right\rangle,
\end{equation}
where the delta function is used to fix the energy level in the middle
of the band. The average of eigenvalues can be performed using the
appropriate two-level cluster function.\cite{mehta} In the limit
$N\rightarrow \infty$, we find
\begin{equation}
\label{eq:Delta2GSE}
\left. \frac{1}{\Delta^2} \right|_{\beta=4} = \frac{\pi^2}{9\,
\delta^2},
\end{equation}
with $\delta$ denoting the inverse of the mean density of
states.\cite{obs3} The ensemble average of off-diagonal matrix
elements of the magnetization is also easily computed in terms of the
trace of the magnetization operator in the limit of large $N$:
\begin{equation}
\label{eq:M2}
\left\langle |\langle n\sigma;0 |\hat{M}| n^\prime\sigma^\prime;0
\rangle |^2 \right\rangle_{n\neq n^\prime,\, \beta=4} = \frac{{\rm Tr}
\left( \hat{M}^2 \right)}{4\,N^2},
\end{equation}
and
\begin{equation}
\label{eq:M4}
\left\langle |\langle n\sigma;0 |\hat{M}| n^\prime \sigma^\prime;0
\rangle |^4 \right \rangle_{n\neq m,\, \beta=4} = \frac{1}{8}\, \left[
\frac{{\rm Tr} \left( \hat{M}^2 \right)}{N^2} \right]^2.
\end{equation}
Inserting Eqs. (\ref{eq:Delta2GSE})--(\ref{eq:M4}) into
(\ref{eq:K2GSE}) we arrive at
\begin{equation}
\label{eq:K2M}
\left\langle K_n^2 \right\rangle_{\beta=4} = \frac{\pi^2}{6\, \delta}
\left[ \frac{{\rm Tr} \left( \hat{M}^2 \right)}{N^2} \right]^2.
\end{equation}
For the symplectic ensemble, one can show\cite{fyodorov95,vonoppen95}
that $\sqrt{\langle K_n^2 \rangle} = (\pi\sqrt{3}/4) \langle |K_n|
\rangle$. Thus, using this relation and combining Eqs. (\ref{eq:g2M})
and (\ref{eq:K2M}), we obtain Eq. (\ref{eq:gK}).

Equation (\ref{eq:gK}), like Eq. (\ref{eq:g2}), also involves only
quantities that are directly measurable. However, in practice, the
difficulty in obtaining large statistics for the second derivative at
$B=0$ from the tunneling conductance data makes it less appealing when
applied to experiments.\cite{petta01,petta02}


Once the variance of the level velocity is obtained from the
experimental data, it may also allow for another test of RMT. Consider
the level velocity correlation function\cite{szafer93,simons93}
\begin{eqnarray}
C_{\mu}(\Delta B) & = & \frac{1}{\delta^2} \left[ \left\langle \left(
\frac{d\varepsilon_\nu}{dB} \right)_{B=B_0+\Delta B}\, \left(
\frac{d\varepsilon_\nu}{dB} \right)_{B=B_0} \right\rangle
\right. \nonumber \\ & & \left. - \left\langle \left(
\frac{d\varepsilon_\nu} {dB} \right)_{B=B_0} \right\rangle^2 \right].
\end{eqnarray}
For a pure ensemble, this correlation function can be rescaled to a
universal form. Defining the correlation field as
\begin{equation}
\label{eq:BcC0}
B_c \equiv 1/\sqrt{C_\nu(0)}
\end{equation}
and calling $x = \Delta B/B_c$ and $c(x) = B_c^2\, C_\nu(\Delta B)$,
the dimensionless correlation function in the unitary ensemble has the
asymptotes\cite{simons93b}
\begin{equation}
c(x) = \left\{ \begin{array}{cc} 1 - 2 \pi^2 x^2, & x \ll 1, \\
-1/(\pi\, x)^2, & x \gg 1. \end{array} \right.
\end{equation}
The full shape of the correlation function is known from numerical
simulations,\cite{szafer93,simons93b} as well as from analytical
calculations.\cite{smolyarenko03}

Finally, it is interesting to note that the correlation field is
related to the amount of disorder in the grains when the electron
motion is diffusive. It is straightforward to show that at $\tau_{\rm
so} \rightarrow \infty$,
\begin{equation}
\label{eq:Bc}
B_c = \kappa\, \frac{\Phi_0/L^2}{\sqrt{k_F^2\, l\, L}},
\end{equation}
where $\Phi_0$ is the flux quantum, $L$ is the grain linear size,
$k_F$ is the Fermi wavelength, and $\kappa$ is a dimensionless
coefficient that depends on the grain geometry. For a spherical shape,
$\kappa = 3\pi/(2\sqrt{5})$, in which case $L$ is the grain radius. By
measuring the variance of the level velocity at large fields, one can
obtain $C_\nu(0)$ and find the experimental value of $B_c$ from
Eq. (\ref{eq:BcC0}). Using Eq. (\ref{eq:Bc}), one can then get an
independent estimate of the amount of disorder present in the grain.
Another approach is to fit the universal
curve\cite{szafer93,simons93b} $c(x)$ to the experimental data and
obtain $C_\nu(0)$ as a fitting parameter.


In summary, we have shown that it is possible to relate the second
moment of the $g$ factor of metallic nanoparticles with strong
spin-orbit coupling to other spectral statistics of energy levels {\it
without} resorting to any microscopic parameter. Our results also show
that it is possible to estimate the spin and orbital contributions to
the fluctuations of the $g$ factor by comparing data taken from
nanoparticles doped and undoped with a heavy-element metal. We suggest
that a fitting of the data to a universal, dimensionless level
velocity correlation function may provide an additional test of the
applicability of random matrix theory to these systems and allow us to
extract information about the intragrain disorder.


This work was supported in part by NSF grants No. DMR 02-37296,
No. DMR 04-39026 (L.I.G.) and No. CCF 0523603
(E.R.M.). C.H.L. acknowledges partial support in Brazil from CNPq,
Instituto do Mil\^enio de Nanoci\^encias, and
FAPERJ. E.R.M. acknowledges partial support from the Interdisciplinary
Information Science and Technology Laboratory (I$^2$Lab) at UCF. We
are grateful to Y. Fyodorov, J. Petta, and F. von Oppen for
enlightening discussions. L.I.G. and E.R.M. thank the Instituto de
F\a'{\i}sica at UERJ, Brazil, and the Aspen Center for Physics for the
hospitality.




\begin{thebibliography}{33}
\bibitem{obs2} The magnetic moment operator is antisymmetric with
respect to time reversal. Therefore, eigenvectors related by time
reversal have eigenvalues equal in amplitude but with opposite signs.
\bibitem{obs3} At $B=0$, due to Kramers degeneracy, $\delta$ is equal
to half of the mean level spacing.
\end{thebibliography}
\end{document}